\documentclass[submission,copyright,creativecommons]{eptcs}
\providecommand{\event}{TTC 2011} % Name of the event you are submitting to

\usepackage{xcolor}
\usepackage{latexsym}
\usepackage{dsfont}
\usepackage{amsmath}
\usepackage{amsfonts}
\usepackage{amssymb}
\usepackage{xspace}
\usepackage{url}

\usepackage{graphicx}

\usepackage{wrapfig}
\usepackage{subfigure}

\usepackage{listings}

\newcommand{\henshin}{\textsc{Henshin}\xspace}

\begin{document}

\title{Saying Hello World with \henshin\xspace - \\A Solution to the TTC 2011 Instructive Case}

\author{Stefan Jurack
\institute{Universit\"at Marburg, Germany}
\email{sjurack@mathematik.uni-marburg.de}
\and Johannes Tietje
\institute{Technische Hochschule Mittelhessen, Gie{\ss}en, Germany}
\email{johannes.tietje@mni.th-mittelhessen.de}}

\def\titlerunning{Saying Hello World with Henshin}
\def\authorrunning{S. Jurack \& J. Tietje}

\maketitle

\begin{abstract}
This paper gives an overview of the \henshin solution to the \emph{Hello World} case study of the Transformation Tool Contest 2011, intended to show basic language concepts and constructs.
\end{abstract}
% ##########################################################################

\section{Introduction}
\label{sec:introduction}
In the modeling community, the Eclipse Modeling Framework (EMF) \cite{emf} has evolved to a well-known and widely used technology.
It is obviously promising to provide model transformation support for EMF.
With \henshin \cite{ABJKT10,Henshin}, a declarative transformation language and tool environment for EMF models is available.
Particularly, EMF models are transformed directly, called \emph{in-place}, without the need of back and forth conversions and copies.
\henshin supports the transformation of EMF models by generated code as well as those created dynamically.
Its transformation concepts base on the well-founded theory of algebraic graph transformation with pattern-based rules as main artifacts, extended by nestable application conditions and attribute computation.
Moreover, the \henshin transformation language allows to structure rules by means of nestable transformation units with well-defined operational semantics.
Additionally, a state space generator supports reasoning by model checking which is, however, not further investigated here.
The export of rules to AGG \cite{AGG} is possible as well, where analysis concerning conflicts and dependencies as well as termination may take place.
Graphical editors support the definition of \henshin model transformations in several fashions.

In the following, Sec.~\ref{sec:henshin} outlines \henshin's transformation language and concepts. A representative subset of the \emph{HelloWorld} solution is presented in Sec.~\ref{sec:solution} followed by a conclusion.

% ##########################################################################

\section{\henshin Transformation in a Nutshell}
\label{sec:henshin}
The \henshin transformation meta-model is an EMF model itself.
Its main artifacts are rules essentially consisting of two graphs, a left-hand side (LHS) and a right-hand side (RHS), specifying model patterns on abstract syntax level.
The LHS describes the pattern to be found while the RHS describes the resulting pattern.
Nodes and edges occurring in the LHS or RHS only, are deleted or created, respectively.
Node mappings between LHS and RHS declare identity, i.e., related nodes and edges are preserved.
Rules may also be equipped with positive and negative application conditions (PACs and NACs) being graph patterns again which confine the match furthermore.
Their true power emerges in combination with logical formulas (AND, OR, NOT) and with nesting even allowing to define conditions over conditions.
Moreover, in the context of rules attribute conditions can be defined whose expressions are evaluated by a JavaScript engine at runtime.

The order of rule applications can be controlled by transformation units, short \emph{units}, providing a well-defined operational semantics.
Units may contain other units including rules which can be considered as atomic units corresponding to their single application.
\henshin offers a predefined set of units: 
\emph{IndependentUnits} provide a non-deterministic unit choice, \emph{PriorityUnits} specify priorities in which contained units are chosen to be applied, a counted application is provided by the \emph{CountedUnit} at which a \emph{count} value of \emph{-1} specifies a loop, i.e., ``as often as possible'', \emph{SequentialUnits} apply rules in a sequential order and perform a rollback if any of its rules cannot be applied, and \emph{ConditionalUnits} allow to specify an \emph{if} condition with a corresponding \emph{then} and \emph{else} part.
Last but not least, \emph{AmalgamationUnits} are special units expressing forall-operations on recurring model patterns.
Here, a \emph{kernel} rule is applied once with a number of \emph{multi} rules being applied as often as possible taking the match of the kernel rule into account.

The object flow can be specified along the control flow using parameters and parameter mappings.
They may pass objects and values from one unit to another one in order to pre-define (partial) matches.

Currently three different editors provide three different views on \henshin transformation models.
A \emph{tree-based editor} provides a very low-level view on the internal model structure, while two other editors offer a more sophisticated graph-like visualization.
One visual editor, in the following called \emph{complex editor}, shows LHS, RHS and application conditions in separate views and is particularly suitable for complex transformation systems with arbitrary control and object flows.
In contrast, the visual \emph{integrated-rule editor} shows rules in an integrated manner in a single view utilizing stereotypes to denote creation, deletion and preservation.
This editor is more suitable for simple rules.
Note, the figures in this paper show one or the other visual representation according to its appropriateness in each individual case.

The application of rules and transformation units on arbitrary models can be triggered by a wizard on the one hand.
On the other hand, \henshin has been designed in a modular way with an independent transformation engine such that it can be freely integrated into any Java project relying on EMF models.
Flexible properties allow to switch between injective (default) and non-injective matching.
Dedicated classes, \texttt{RuleApplication} and \texttt{UnitApplication}, provide appropriate methods to select and apply rules and transformation units, respectively.

% ##########################################################################

\section{Case Study Tasks Solved by \henshin}
\label{sec:solution}

In the following a representative subset of the complete solution of the \emph{Hello World} challenge \cite{helloworldcase} is presented while a full listing is given in Appendix~\ref{apx:AllSolutions}.
Each transformation is triggered via Java source code. In order to display/persist a transformation's outcome, appropriate rules or units contain a parameter, e.g., \texttt{result} or \texttt{counter}, which is fetched on code level (see Appendix~\ref{apx:codeListing}).
%
%Please note, that each rule or unit contains at least one parameter, e.g., \texttt{result} or \texttt{counter}, used to return and display/persist the outcome on code level (see Appendix~\ref{apx:codeListing}).
Furthermore, in some cases we exploit a \henshin self-contained helper structure called \emph{trace model} in order to keep track of the transformation process.
This model consists of a class \texttt{Trace} which might be connected to any EMF object by two references, \texttt{source} and \texttt{target}.

%In the following we present a representative subset of our complete solutions to all the different \emph{Hello World} case study tasks.
%As we have several editors available to show the rules we choose the one that suits best in each individual case.
%Please note, that each rule or unit contains a parameter, e.g., \emph{result} or \emph{counter}, used to return and display/persist the outcome on code level (see Appendix~\ref{apx:codeListing}).
%Note furthermore, that in some cases we exploit a \henshin self-contained helper structure called \emph{trace model} in order to keep track of the process of a transformation.
%This model consists of class \emph{Trace} which might be connected to any EMF object by two references, source and target.
%%In order to keep track of the transformations in subsections \ref{subsec:count} to \ref{subsec:transitive}, we use the henshin-own fully generic Trace model instead of a self-provided mapping model to mark model elements which are already processed.

%##################################################################################

\begin{figure}[htbp]
	\centering
		\includegraphics[width=1.00\textwidth]{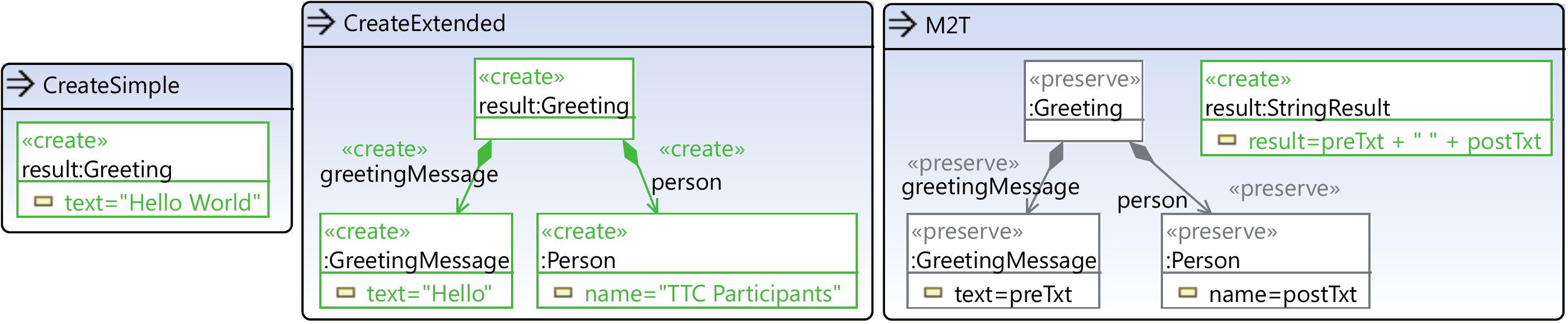}
	\caption{Tree rules concerning the \emph{Hello World!} subtasks shown in an integrated manner.}
	\label{fig:helloWorld}
\end{figure}

\paragraph{Task 1: Hello World!}
\label{solution:HelloWorld}
Figure~\ref{fig:helloWorld} shows the solutions for each of the tree subtasks of the first \emph{Hello World!} task by means of the integrated-rule editor.
Each rule is presented as rounded rectangle with its name at the top and its graph structure contained.
Stereotypes denote related nodes and edges to be \texttt{create}d, \texttt{delete}d or \texttt{preserve}d. 
Note, in the editor undefined stereotypes default to \texttt{preserve}. However, it is conceivable to omit that stereotype at all in the future.

In the left of Fig.~\ref{fig:helloWorld}, rule \texttt{CreateSimple} consists of a node of type \texttt{Greeting} to be created which is assigned to the parameter \texttt{result}.
The value of its attribute \texttt{text} is also set to the string \texttt{"Hello World"}.
Rule \texttt{CreatedExtended}, located in the center of Fig.~\ref{fig:helloWorld}, creates a more complex greetings structure by introducing a number of nodes and edges.
The model-to-text transformation subtask is represented by the rule \texttt{M2T} which preserves a given greetings structure while creating a new \texttt{:StringResult} object.
In addition, two parameters, \texttt{preTxt} and \texttt{postTxt}, are utilized to fetch the corresponding attribute values of the matched objects \texttt{:GreetingMessage} and \texttt{:Person} in order to participate in the attribute computation for the \texttt{:StringResult} object.

%Figure~\ref{fig:helloWorld} shows the solution for the simple and the extended \emph{Hello World} tasks by means of our integrated-rule editor.
%Each rounded rectangle represents a rule with its name at the top and its content below.
%Rule \emph{CreateSimple} contains a node of type \texttt{Greeting} assigned to parameter \emph{result}.
%Nodes may have attributes with values assigned, e.g., \texttt{text="Hello World"}.
%Stereotype \texttt{<<create>>} denotes that this node is to be created which is equivalent to this node being in an RHS without any mappings to the LHS.
%Rule \emph{CreatedExtended} creates a more complex greeting structure also introducing new edges and rule \emph{M2T} performs a model-to-text transformation in the sense that the greeting structure remains unchanged while a new \texttt{StringResult} object is created.
%Note that we use additional parameters here, \texttt{preTxt} and \texttt{postTxt}, which are filled with its related values during matching and which are then used to calculate a new attribute value.
%

%##################################################################################

\begin{figure}[htbp]
	\centering
		\includegraphics[width=0.80\textwidth]{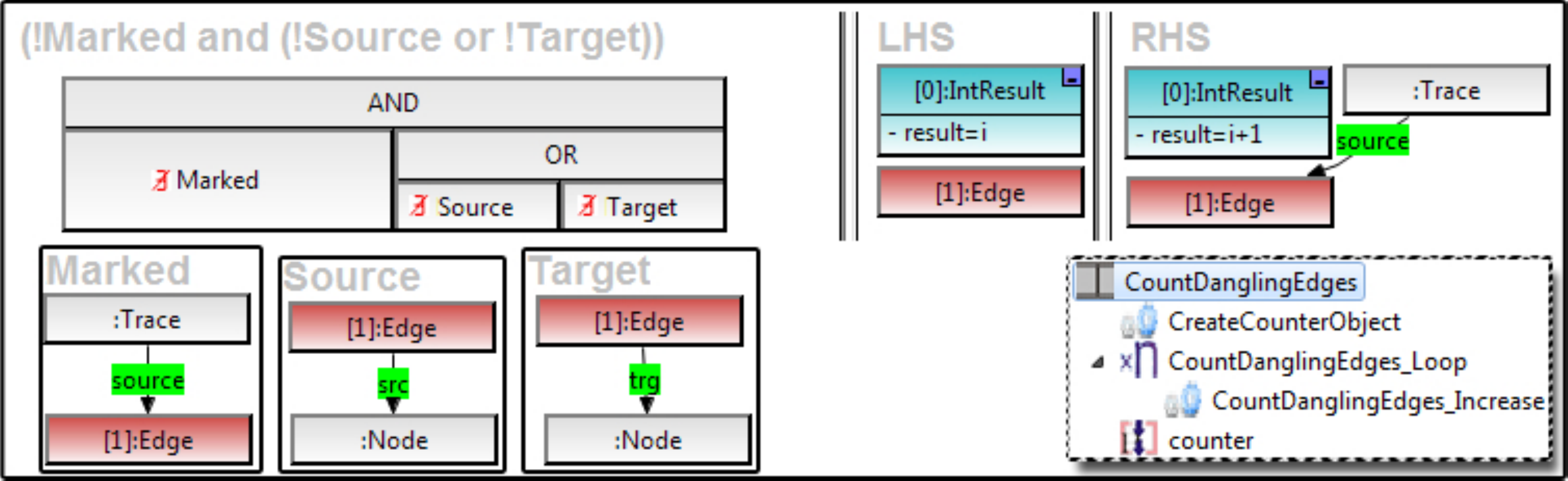}
	\caption{Overview of the sequential unit \texttt{CountDanglingEdges} (bottom right) and details of the rule \texttt{CountDanglingEdges\_Increase} with LHS, RHS and three NACs composed by a logic formula.}
	\label{fig:CountDanglingEdges}
\end{figure}

\paragraph{Task 2: Count Matches.}
\label{solution:CountMatches}
All related subtasks are implemented by structurally similar \emph{sequential units} while one, \emph{CountDanglingEdges}, is exemplarily illustrated in Fig.~\ref{fig:CountDanglingEdges} by means of the complex editor.
Actually, this figure is a compositions of different views of the complex editor in favor of space savings. 
The control flow is shown in the lower right where at first rule \texttt{CreateCounterObject} is applied.
It creates an initial \texttt{counter:IntResult} object used for counting and used as return value with the help of parameter \texttt{counter}.
Note that the \emph{same} rule is \emph{reused} by all subtasks.
Afterwards, contained in a counted unit \texttt{CountDanglingEdges\_Loop}, rule \texttt{CountDanglingEdges\_Increase} is applied as often as possible performing the actual counting.
Its LHS, RHS (upper right) and NACs (left) are shown in Fig.~\ref{fig:CountDanglingEdges} as well.
Identical nodes are indicated by numbers in brackets.
This rule matches the \texttt{:IntResult} object and an \texttt{:Edge} object which has not been marked by a \texttt{:Trace} object yet \emph{and} which has no source \emph{or} no target, including not having either.
Each rule application increases the value of the attribute \texttt{result} by one and marks the edge as being visited by creating and connecting a \texttt{:Trace} object.
%
%Each transformation unit contains at first the rule \emph{CreateCounterObject} which creates a foundational \texttt{IntResult} object for the following rule application, with an initial value of ``0''.
%
%Besides this basic rule, each transformation unit contains a counted unit which applies its respective child rule (\emph{CountLoopingEdges}, \emph{CountNodes}, \emph{CountIsolatedNodes}, \emph{CountCircles} and \emph{CountDanglingEdges}) as often as it is possible to match all model elements. Each child rule increases the counter by ``1'' for each application and marks the matched elements with a \texttt{Trace} object to exclude them of further examination.

\begin{figure}[htbp]
	\centering
		\includegraphics[width=0.80\textwidth]{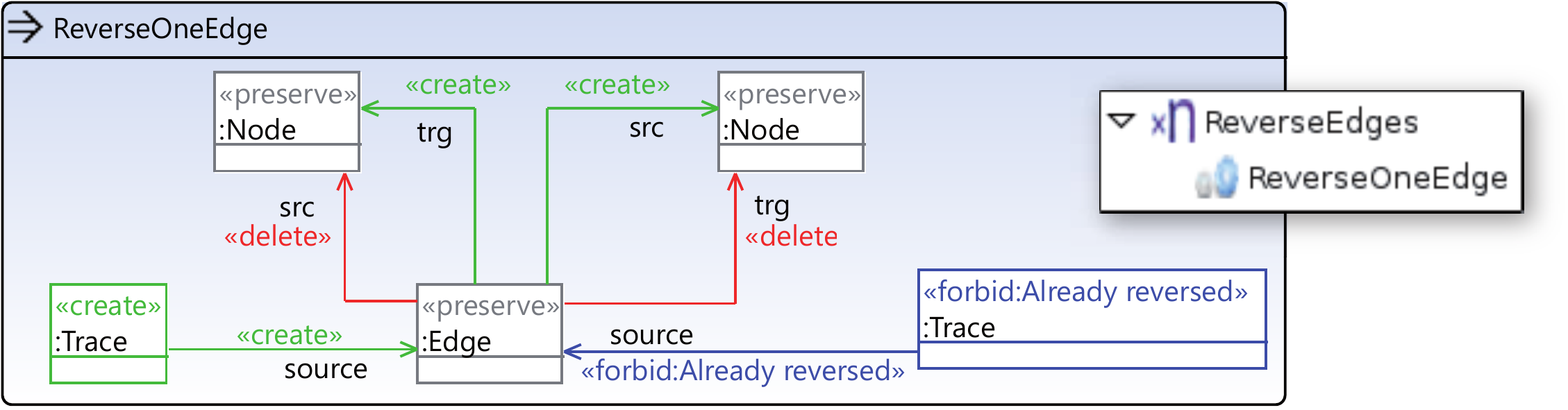}
	\caption{Rule \texttt{ReverseOneEdge} for reversing an edge's direction.}
	\label{fig:ReverseEdges}
\end{figure}

\paragraph{Task 3: Reverse Edges.}
\label{solution:ReverseEdge}
This task is solved by applying the rule \texttt{ReverseOneEdge} (cf. Fig.~\ref{fig:ReverseEdges}) in a loop as often as possible.
\texttt{:Edge} and \texttt{:Node} objects are preserved during the transformation while two references are \texttt{create}d and two others are \texttt{delete}d.
In addition, a \texttt{:Trace} object is created as marker in combination with the NAC ``Already reversed'' \texttt which {forbid}s such a marker to exist in the given context.

Another solution would have been using an amalgamation unit.
In this case, the kernel rule would be empty.
The multi rule would be similar to the rule \texttt{ReverseOneEdge} but without the NAC and the trace objects, since a single application of an amalgamation unit matches each multi rule as often as possible and performs all those transformations in parallel. See Task 5 for more details on amalgamation units.

%This task is solved by a counted unit (count=-1) applying rule \emph{ReverseOneEdge} (cf. Fig.~\ref{fig:ReverseEdges}) as often as possible.
%\texttt{Edge}s and \texttt{Node}s are preserved during the transformation while the references are not: 
%Two are \emph{create}d and two others are \emph{delete}d.
%In addition, a \texttt{Trace} object is used as marker again in combination with a negative application condition \emph{forbid}ding such a marker to exist already in the given context.
%
%The \emph{Update} task contains a counted unit which applies its child rule \emph{ReverseOneEdge} as often as possible, which reverses a single edge (i.e., swaps the values of its \texttt{src} and \texttt{trg} references) and marks it with a \texttt{Trace} object.

\paragraph{Task 4: Simple Migration.}
\label{solution:SimpleMigration}
The solution for this task uses techniques already known.
At first the rule \texttt{CreateNewGraph} creates a new \texttt{Graph} object as initial root object. 
%Sequentially applied, at first the rule \texttt{CreateNewGraph} creates a new \texttt{Graph} object as initial root object. 
Afterwards, one rule (\texttt{MigrateNode}) migrating a single node is applied in a loop and subsequently one rule (\texttt{MigrateEdge}) migrating an edge is applied in a loop too.
Again, \texttt{:Trace} objects mark those elements already translated.

There are a number of alternative ways to realize this task.
For example, one could use an amalgamation unit creating a \texttt{:Graph} object in the kernel rule and performing the node migration in a multi rule without the need of trace objects.
Another amalgamation unit could realize the migration of edges.
Then, the whole migration would require the sequential application of both units.

The solution for the extended task is actually realized with the help of an amalgamation unit.
We skip this here and rather refer to the following task using this special kind of unit as well.

%Afterwards two successive counted units follow, one containing the rule dealing with the migration of nodes and the other containing the rule adding corresponding edges.
%Again, \texttt{Trace} objects mark elements already translated and a parameter carries the resulting graph to be returned by the transformation process.

%The solution for the extended task is realized a bit trickier with the help of an amalgamation unit.
%We skip this here and rather refer to the following task using this special kind of unit as well.
%

\begin{figure}[h]
	\centering
		\includegraphics[width=1.0\textwidth]{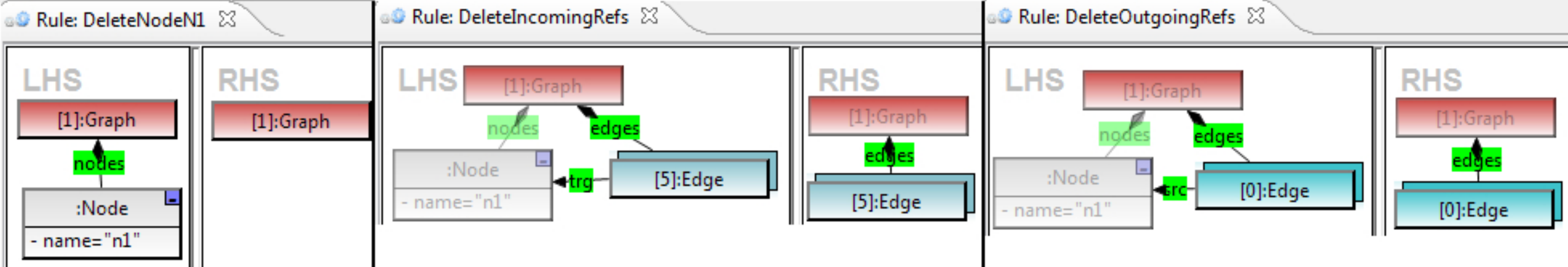}
	\caption{Amalgamation unit \texttt{DeleteNodeN1Simple} consisting of one kernel and two multi rules.}
	%\caption{Kernel rule \texttt{DeleteNodeN1} (left) and two multi rules (center and right) being integral parts of the amalgamation unit \texttt{DeleteNodeN1Simple}.}
	\label{fig:DeleteNodeN1Simple}
\end{figure}

\paragraph{Task 5: Delete Node and its Incident Edges.}
\label{solution:DeleteNode}
Both deletion subtasks have been solved using amalgamation units.
Figure~\ref{fig:DeleteNodeN1Simple} shows all parts of the amalgamation unit \texttt{DeleteNodeN1Simple} realizing the core task.
The left rule, \texttt{DeleteNodeN1}, serves as kernel rule and matches a \texttt{:Graph} object and a connected \texttt{:Node} object named ``n1''.
Its RHS points out that the node is to be deleted.
The rules \texttt{DeleteIncomingRef} and \texttt{DeleteOutgoingRef} serve as multi rules.
Their names indicate already, that they delete an incoming and an outgoing reference, respectively.
Each multi rule is defined in the context of its kernel rule denoted by the parts grayed out in the LHS and RHS of the multi rules, while the parts belonging to the multi rule are illustrated by double rectangles.
An application of the amalgamation unit leads to a single match of the kernel rule and all possible matches of the multi rules taking the match of the kernel rule into account.
For further information on amalgamation units please refer to \cite{BET10}.

The extended task is realized in a similar way and thus not shown here (see Appendix).

\paragraph{Task 6: Insert Transitive Edges.}
\label{solution:TransitiveEdges}
The last task is solved by the rule \texttt{InsertTransitiveEdge} applied in a loop to create a single transitive edge for an already existing connection between three \texttt{Node} objects.
This is quite analogous to what has been explained above and therefore not presented in detail here.

\section{Conclusion}
\label{sec:conclusion}
This paper presents the solution of the \emph{Hello World} challenge \cite{helloworldcase} implemented by \henshin. 
All tasks including all optional tasks are solved.
The implementation is made available under SHARE \cite{ShareHenshin11}.

The solution is particularly characterized by a visual transformation language, pattern-based rules and the capability of defining control and object flows. 
Three different graphical editors support the creation and editing of \henshin transformation systems, each one with its assets and drawbacks.
However, each editor provides a visualization more or less close to the underlying transformation model.
For example, the creation of more than two application conditions leads to quite some nesting due to binary formulas as shown in Fig.~\ref{fig:CountDanglingEdges}.
Moreover, control and object flows are often not easy to grasp.
To overcome these issues a more adequate representation is desirable.
The question arises whether an integrated visualization of rules and control flow (and object flow) or a separate visualization is more appropriate.
Whatever decision will be made, the current major shortcomings of \henshin, understandability and conciseness, need to be addressed in further developments.
The application of each transformation is currently triggered by dedicated source code, although a wizard is available as well.
However, transformation engine configurations, e.g., switch to non-injective matching, require the use of code yet.
Besides the scope of solutions for this task, the performance of \henshin behaves quite well. Additional features to be possibly integrated are attribute computations using OCL expressions, path expressions, and untyped nodes and edges representing elements of arbitrary type improving the understanding of rules.

\bibliographystyle{eptcs}
\bibliography{literatur/literatur}
\newpage

\begin{appendix}

\section{All Solutions}
\label{apx:AllSolutions}

In the following the solutions to all tasks are given.
If appropriate, rules are shown using the integrated-rule editor manually composed with an overview of belonging units.
More complex rules are rather given by means of the complex-rule editor showing LHS, RHS and application conditions separately.

\subsection{Task 1}
\label{apx:task1}
\begin{enumerate}
	\item Constant transformation.
	\item Constant transformation that creates a model with references.
	\item Model-to-text transformation.
\end{enumerate}

\begin{figure}[h!]
\centering
\includegraphics[width=0.25\textwidth]{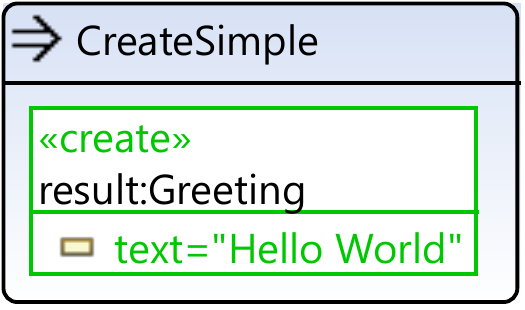}
\caption{Task 1.1 - Rule \texttt{CreateSimple}.}
\end{figure}

\begin{figure}[h!]
\centering
\includegraphics[width=0.5\textwidth]{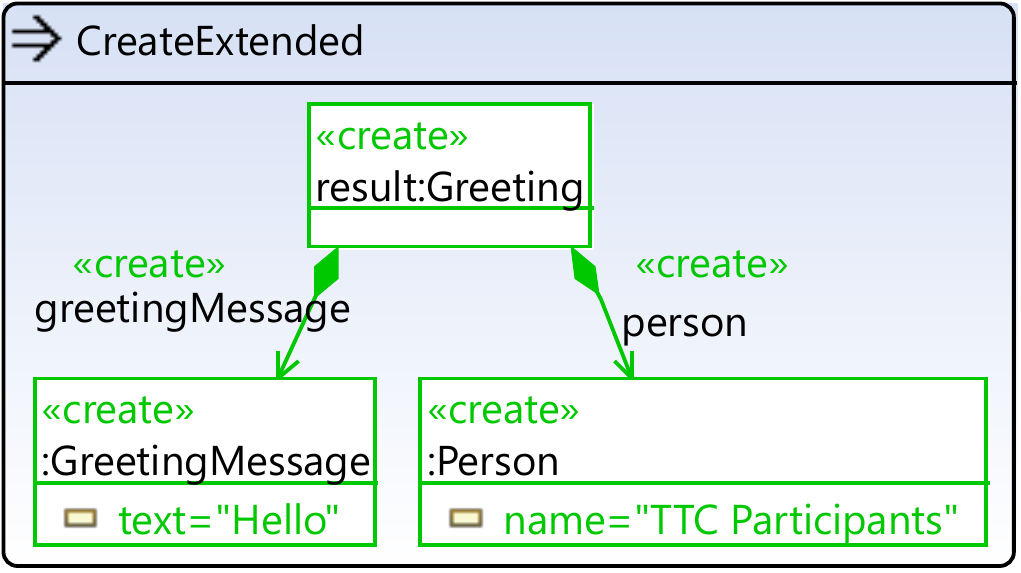}
\caption{Task 1.2 - Rule \texttt{CreateExtended}.}
\end{figure}

\begin{figure}[h!]
\centering
\includegraphics[width=0.6\textwidth]{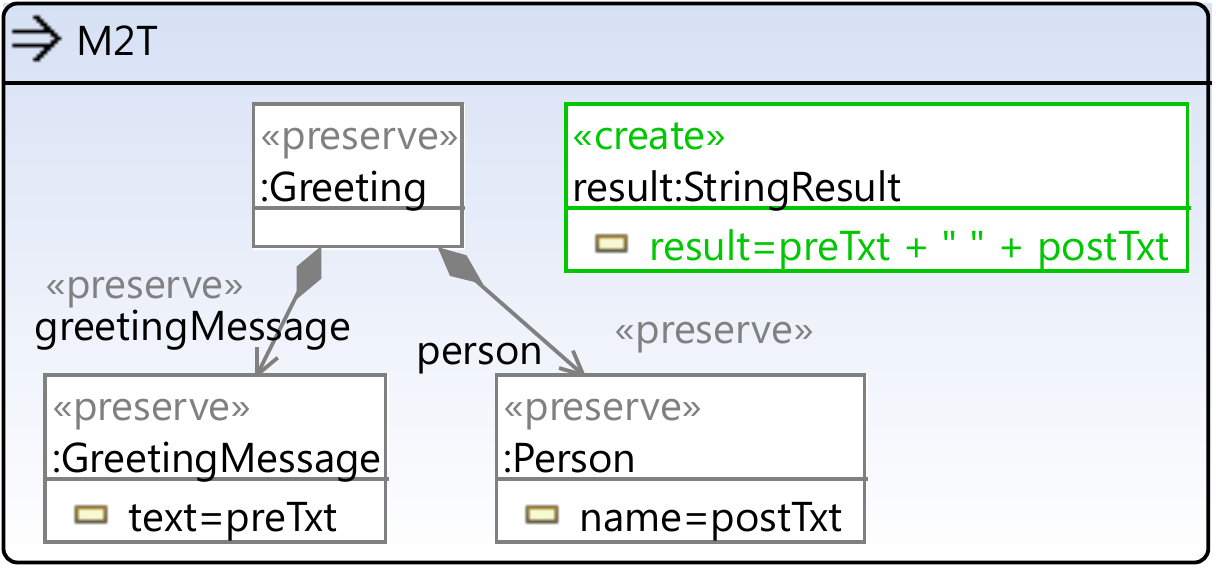}
\caption{Task 1.3 - Rule \texttt{M2T}.}
\end{figure}

\subsection{Task 2} 
\label{apx:task2}
Model query that counts the number of \dots
\begin{enumerate}
	\item nodes in a graph.
	\item looping edges.
	\item isolated nodes.
	\item matches of a circle consisting of three nodes.
	\item dangling edges.
\end{enumerate}
Each task is solved utilizing a sequential unit containing the rule \texttt{CreateCounterObject} at first and subsequently containing a counted unit with another rule contained. 
While all solutions to task 2 reuse the \emph{same} rule \texttt{CreateCounterObject}, the actual counting is performed in the rule contained in the respective counted unit.
The counted units have a \texttt{count} value of \texttt{-1} and thus force the contained rule to be applied in a loop, i.e., as often as possible.
\begin{figure}[h]
\centering
\includegraphics[width=0.25\textwidth]{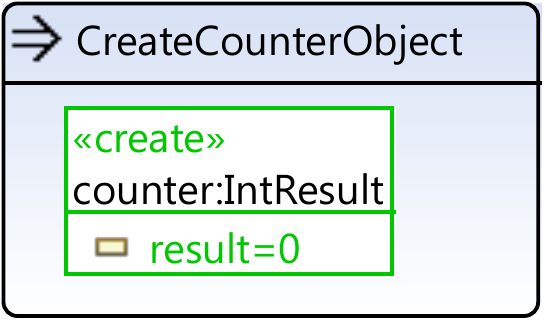}
\caption{Rule \texttt{CreateCounterObject}}
\end{figure}
\begin{figure}
\centering
\includegraphics[width=0.8\textwidth]{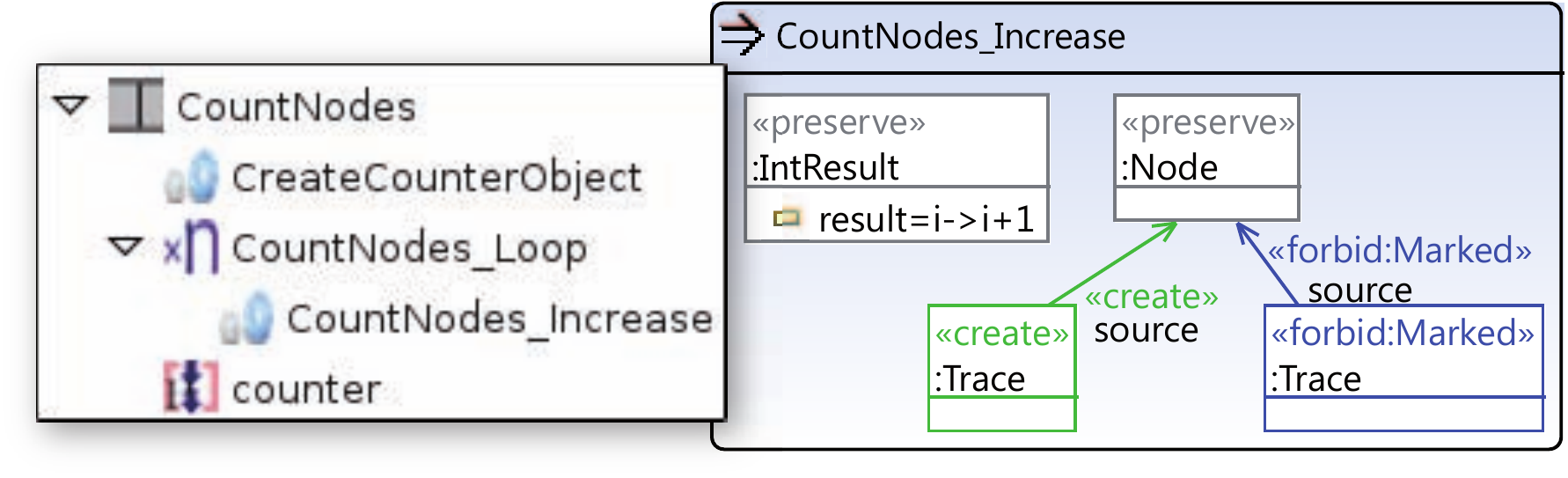}
\caption{Task 2.1 - Sequential unit \texttt{CountNodes} and details of  the rule \texttt{CountNodes\_Increase}.}
\label{apx:fig:task2-1}
\end{figure}
\begin{figure}
\centering
\includegraphics[width=0.69\textwidth]{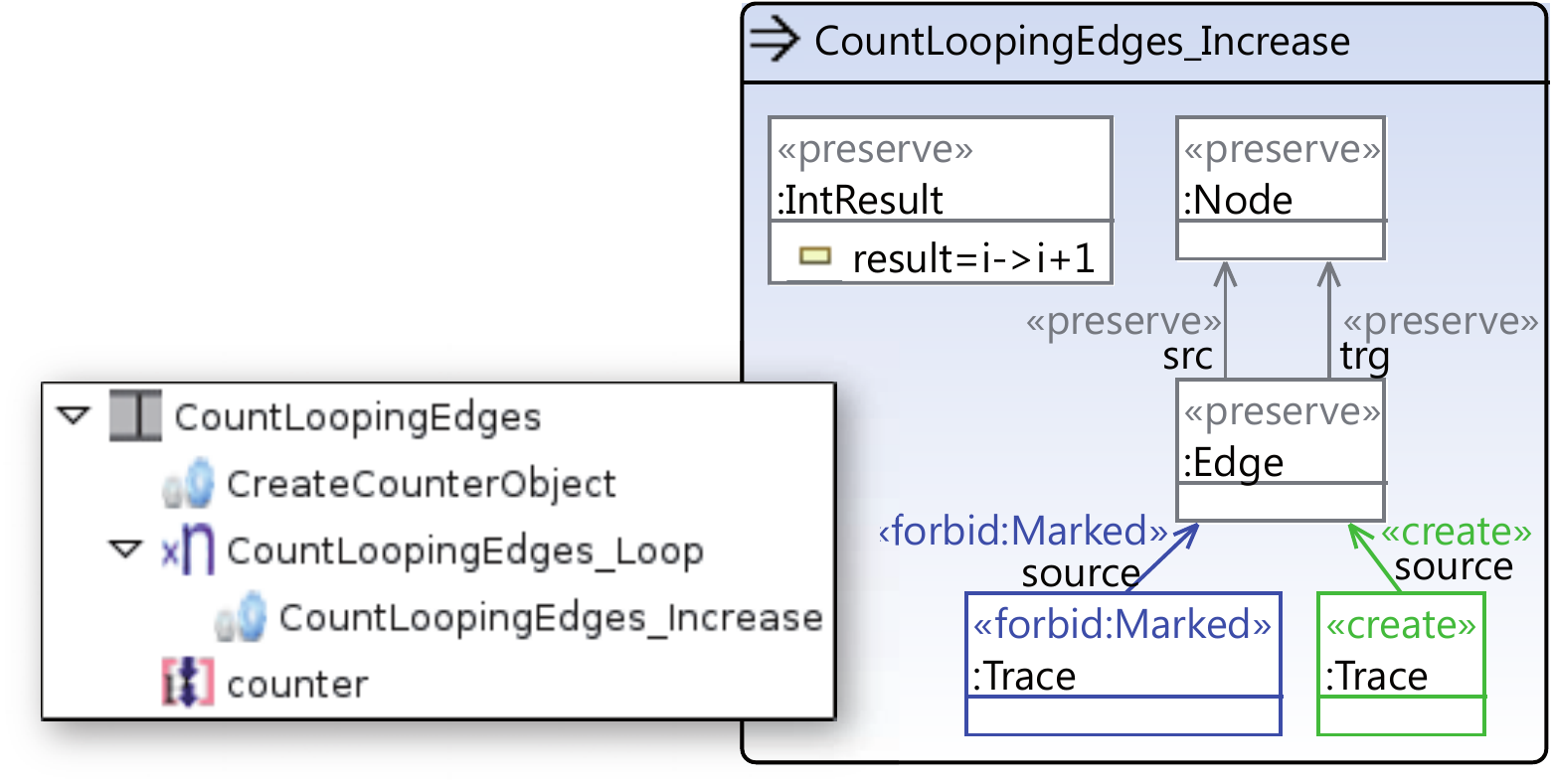}
\caption{Task 2.2 - Sequential unit \texttt{CountLoopingEdges} and details of  the rule \texttt{CountLoopingEdges\_Increase}.}
\label{apx:fig:task2-2}
\end{figure}
\begin{figure}
\centering
\includegraphics[width=0.75\textwidth]{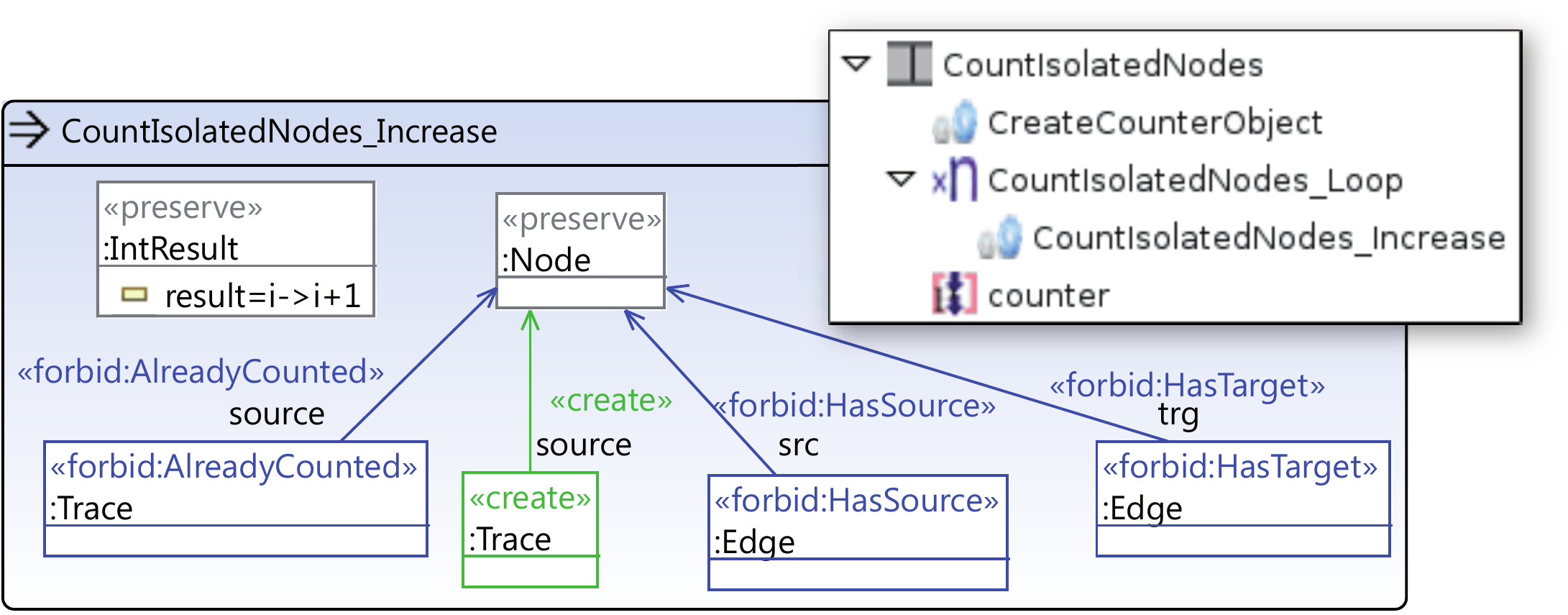}
\caption{Task 2.3 - Sequential unit \texttt{CountIsolatedNodes} and details of  the rule \texttt{CountIsolatedNodes\_Increase}.}
\label{apx:fig:task2-3}
\end{figure}
\begin{figure}
\centering
\includegraphics[width=0.9\textwidth]{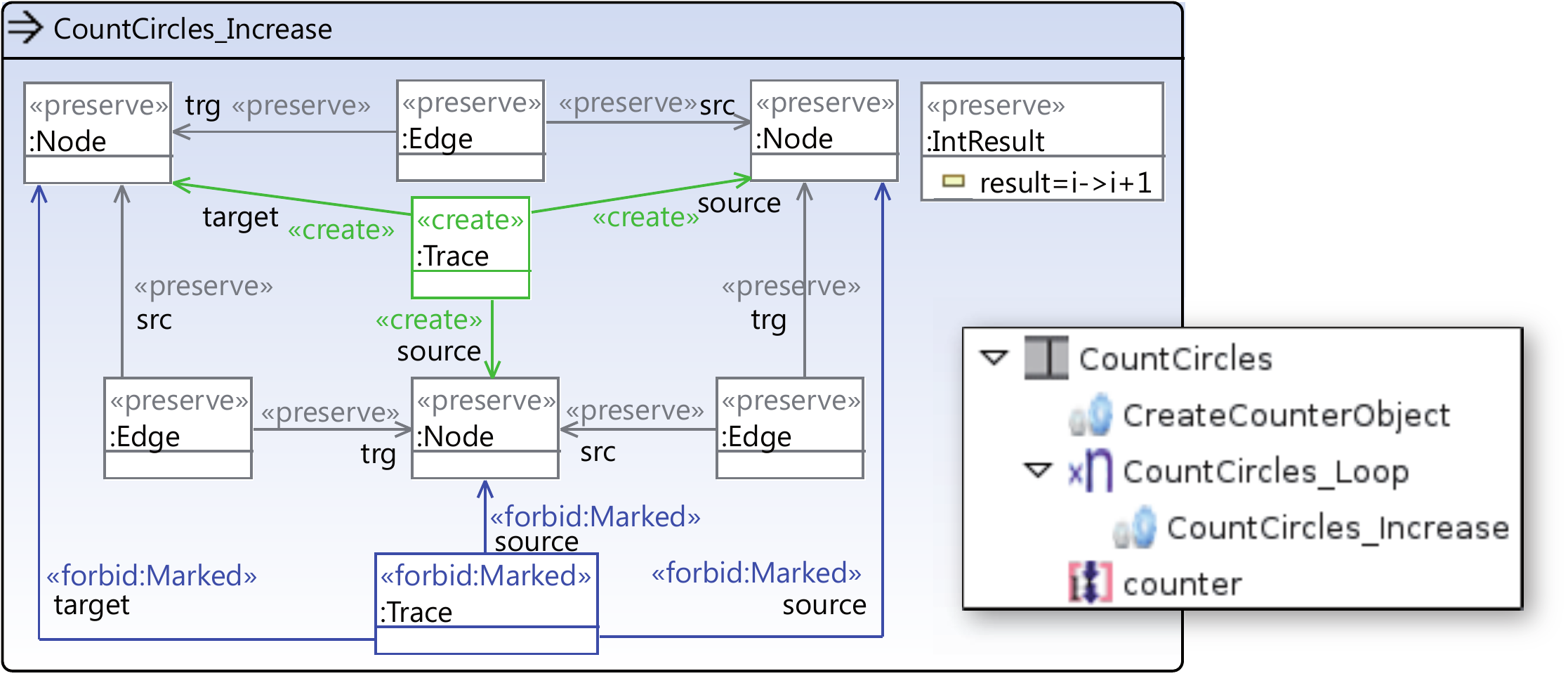}
\caption{Task 2.4 - Sequential unit \texttt{CountCircles} and details of  the rule \texttt{CountCircles\_Increase}.}
\label{apx:fig:task2-4}
\end{figure}
\begin{figure}
	\centering
		\includegraphics[width=0.80\textwidth]{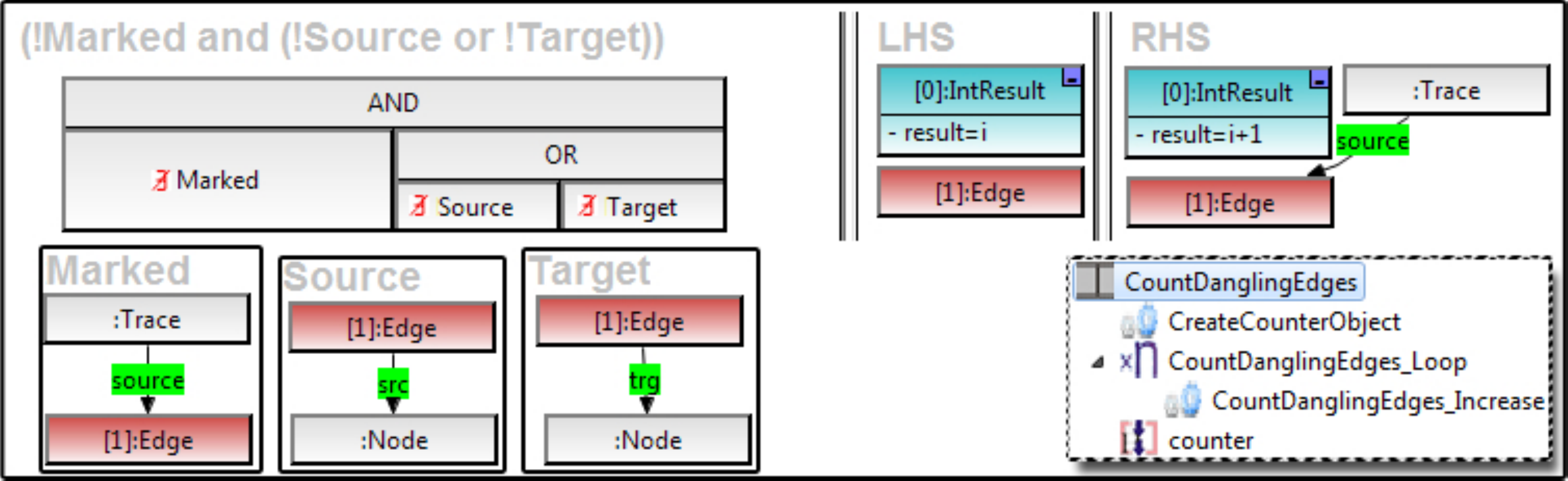}
	\caption{Task 2.5 - Sequential unit \texttt{CountDanglingEdges} and details of the rule \texttt{CountDanglingEdges\_Increase} with LHS, RHS and three NACs composed by a logic formula.}
\label{apx:fig:task2-5}
\end{figure}

\newpage

\subsection{Task 3} 
\label{apx:task3}
\begin{enumerate}
	\item Reversing all edges.
\end{enumerate}
The solution to this task is characterized by a counted unit with \texttt{count=-1} applying rule \texttt{ReverseOneEdge} in a loop, i.e., as often as possible.
\begin{figure}[h]
	\centering
		\includegraphics[width=0.80\textwidth]{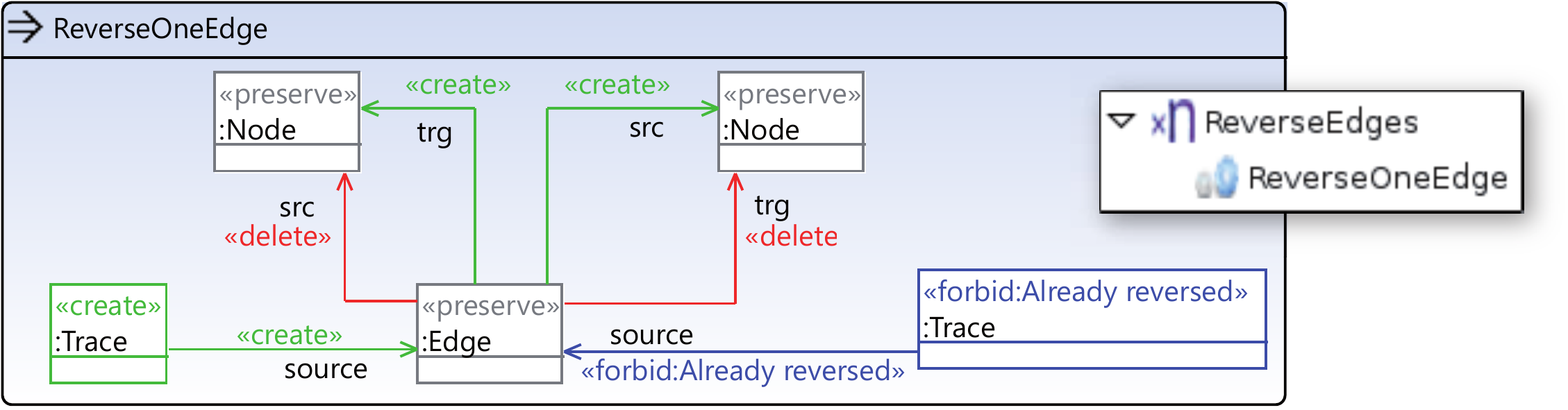}
	\caption{Task 3.1 - Counted unit \texttt{ReverseEdges} and details of the rule \texttt{ReverseOneEdge}.}
\label{apx:fig:task3-1}
\end{figure}

\newpage

\subsection{Task 4} 
\label{apx:task4}
\begin{enumerate}
	\item Simple migration of a graph conforming to one metamodel (\texttt{graph1}) to a graph conforming to a second metamodel (\texttt{graph2}).
	\item Topology-changing migration of a graph conforming to one metamodel (\texttt{graph1}) to a graph conforming to a second metamodel (\texttt{graph3}).
\end{enumerate}
\begin{figure}[h]
	\centering
		\includegraphics[width=0.70\textwidth]{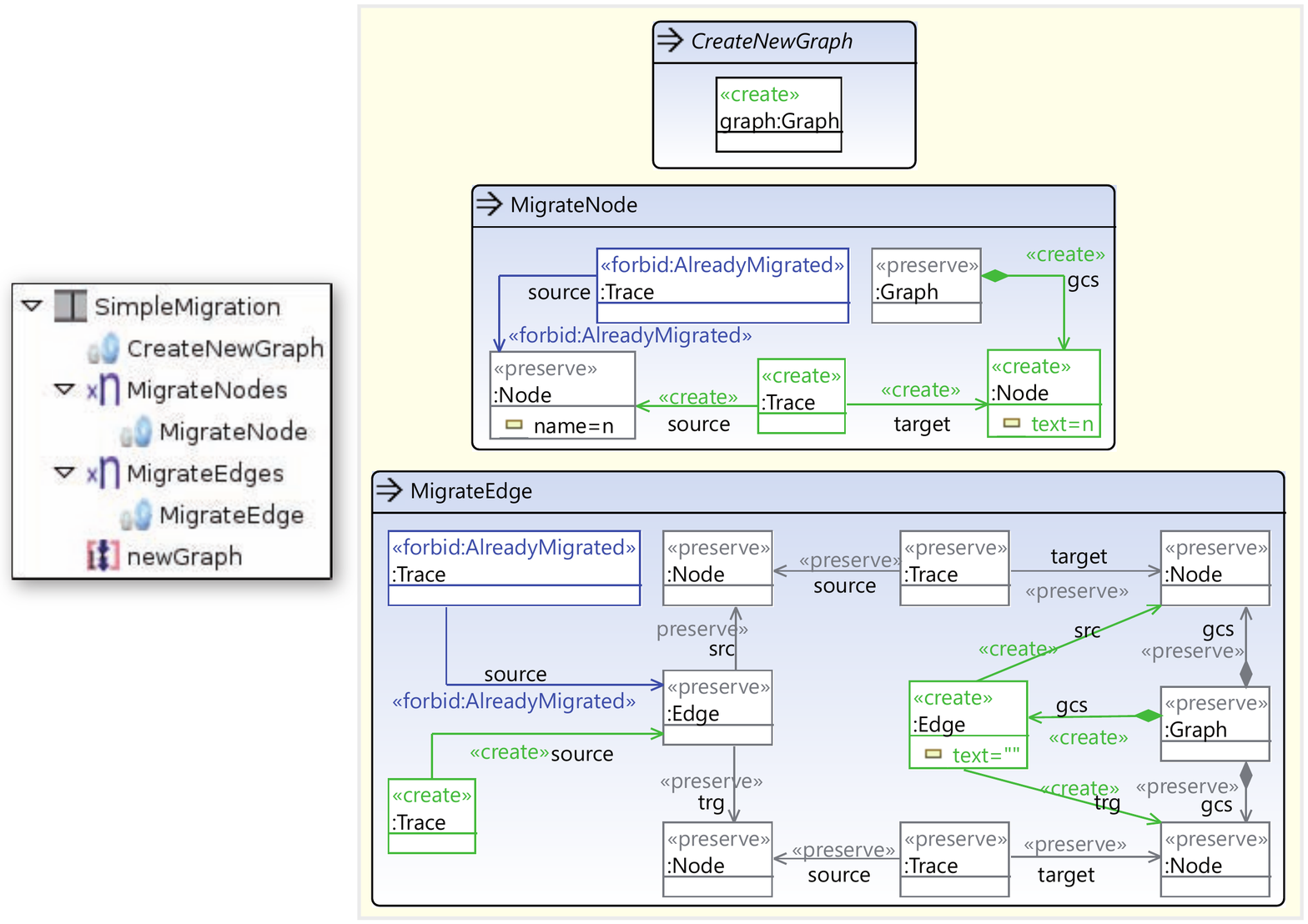}
	\caption{Task 4.1 - Sequential unit \texttt{SimpleMigration} and details of its rules. Within the sequential unit, at first a new \texttt{Graph} object is created serving as container for the migrated nodes and edges. Afterwards, each node is migrated by rule \texttt{MigrateNode} applied in a loop and then each edge is migrated by rule \texttt{MigrateEdge} analogously. Please note a current shortcoming of the visualization of rules: \texttt{MigrateNode} and \texttt{MigrateEdge} show a number of \texttt{Node} and \texttt{Edge} objects which are, however, typed over different metamodels. For example, in \texttt{MigrateNode} the left \texttt{Node} object is typed over the source metamodel (\texttt{graph1}) while the right \texttt{Node} object is typed over the target metamodel (\texttt{graph2}). We plan to overcome this issue by showing the underlying metamodel's name below each node's name, if required.}
\label{apx:fig:task4-1}
\end{figure}
\begin{figure}
	\centering
		\includegraphics[width=1.0\textwidth]{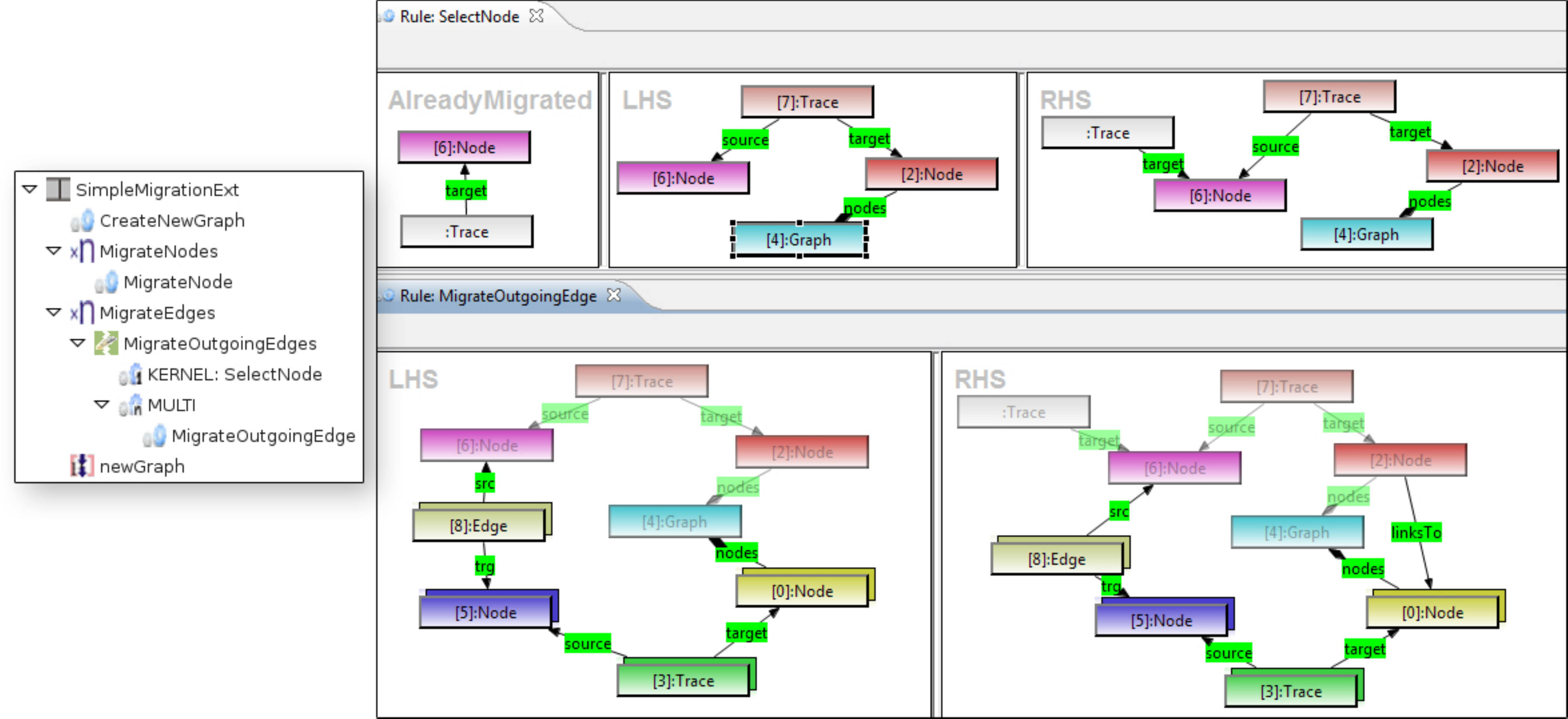}
	\caption{Task 4.2 - Sequential unit \texttt{SimpleMigrationExt} and details of parts of its amalgamation unit. The kernel rule \texttt{SelectNode} matches \emph{one} node and its migrated counterpart node whose edges have not been migrated yet (due to NAC \emph{AlreadyMigrated}) while the multi rule matches \emph{each} related edge and introduces a corresponding \texttt{linksTo} reference in the target model. The RHS of the kernel rule marks the node by a \texttt{Trace} object to be completely migrated including its edges .
		Note, that the rules \texttt{MigrateNode} and \texttt{CreateNewGraph} are structurally equal to corresponding rules of Task 4.1, besides the fact, that they differ in their underlying metamodel. }
\label{apx:fig:task4-2}
\end{figure}

\newpage
\
\subsection{Task 5} 
\label{apx:task5}
\begin{enumerate}
	\item Deletion of a node with name ``n1''.
	\item Deletion of a node with name ``n1'' and all its incident edges.
\end{enumerate}
Each subtask is solved by applying a single amalgamation unit. Note, that the \emph{same} rule \texttt{DeleteNodeN1} is reused by both amalgamation units shown below.
\begin{figure}[h]
	\centering
		\includegraphics[width=0.70\textwidth]{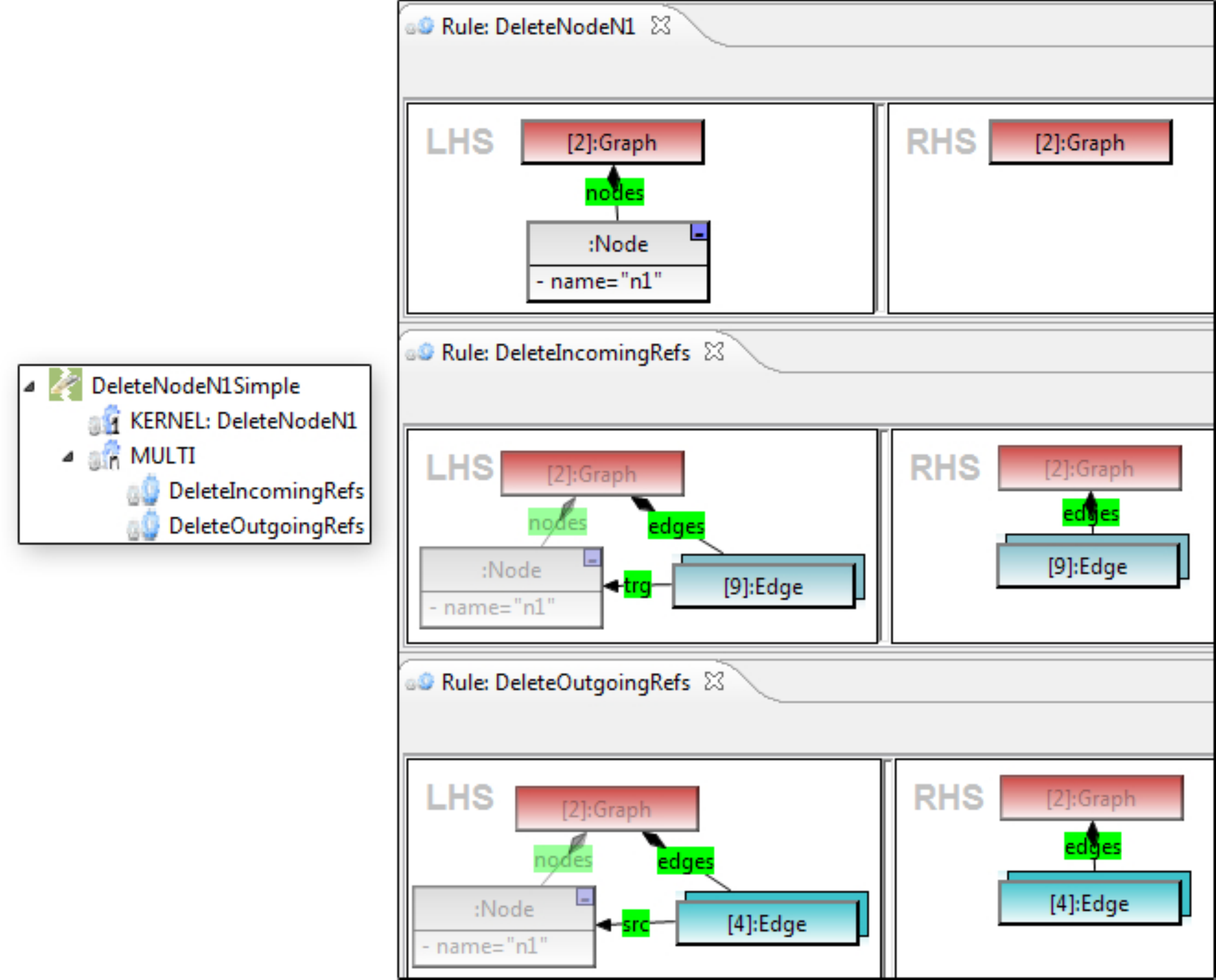}
	\caption{Task 5.1 - Amalgamation unit \texttt{DeleteNodeN1Simple} and details of its rules.}
\label{apx:fig:task5-1}
\end{figure}
\begin{figure}
	\centering
		\includegraphics[width=0.70\textwidth]{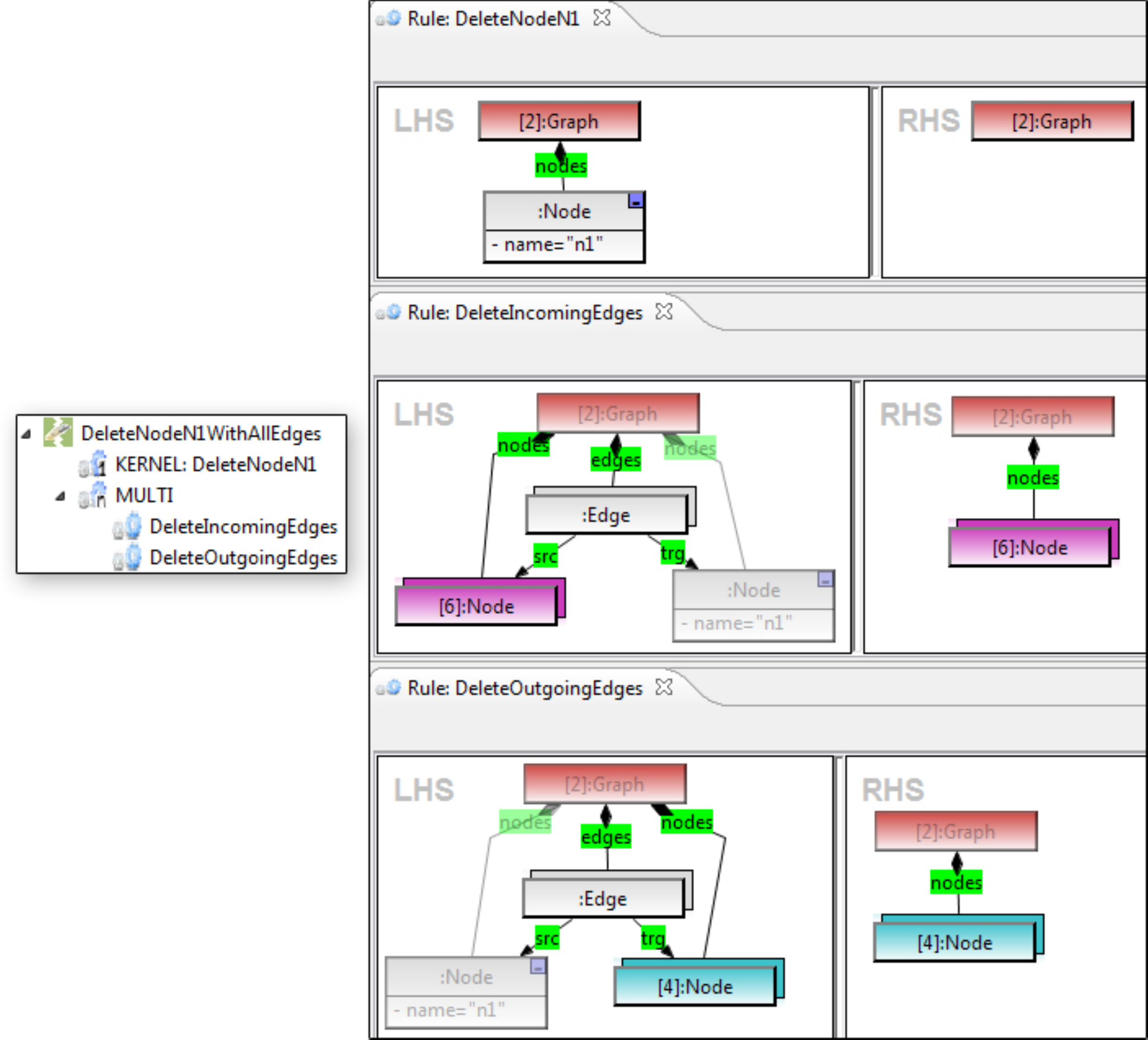}
	\caption{Task 5.2 - Amalgamation unit \texttt{DeleteNodeN1WithAllEdges} and details of its rules.}
\label{apx:fig:task5-2}
\end{figure}

\newpage

\subsection{Task 6} 
\label{apx:task6}
The solution to this task turned out to be quite simple: A single rule needs to be called in a loop only.
\begin{figure}[h]
	\centering
		\includegraphics[width=0.70\textwidth]{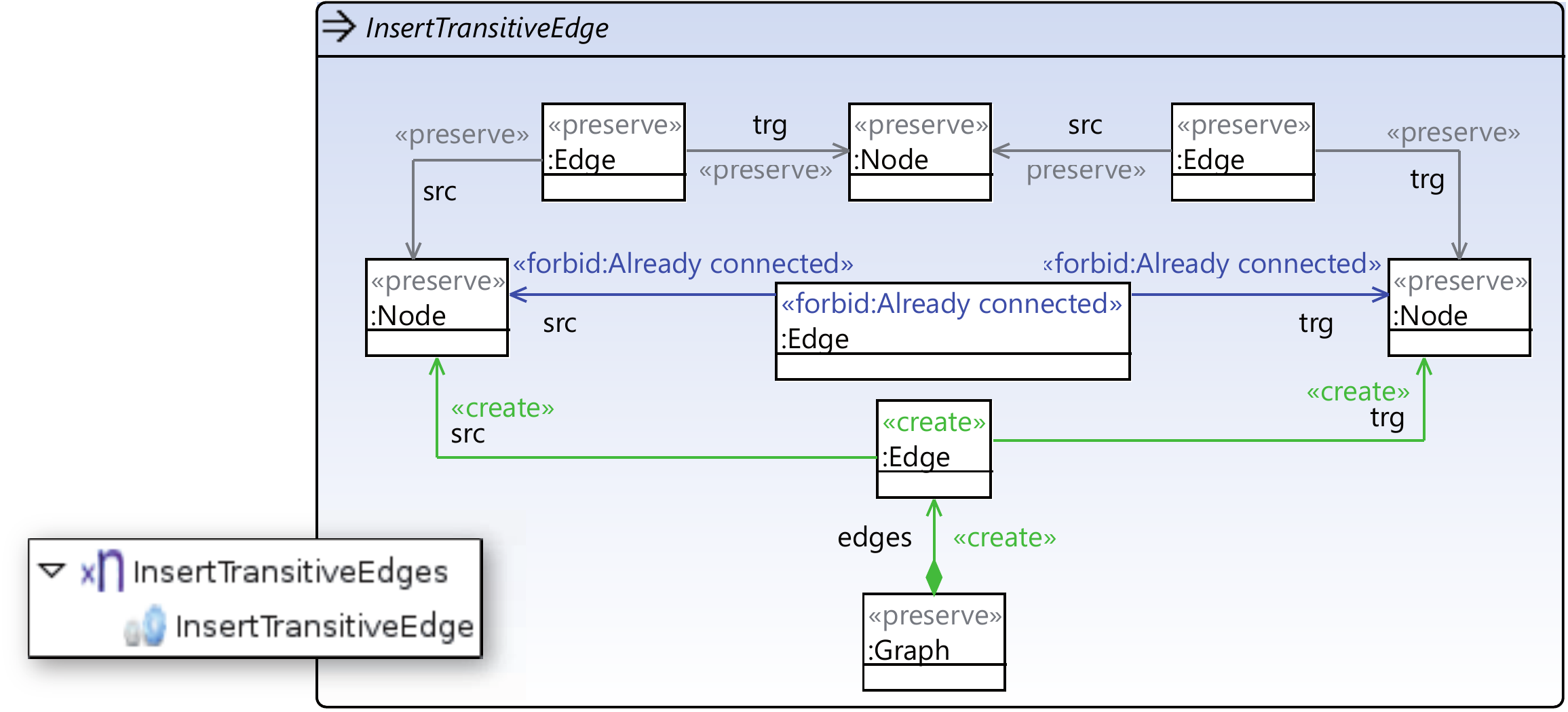}
	\caption{Task 6.1 - Counted unit \texttt{IntertTransitiveEdges} and details of its rule \texttt{InsertTransitiveEdge}.}
\label{apx:fig:task6-1}
\end{figure}

%\section{Overview of the Transformation Algorithms}
%\label{apx:trafoAlgo}
%\input{transformationunits}
%
%\section{Rules}
%\label{apx:rules}
%\input{rules}

\section{Java Code of the Transformation Application}
\label{apx:codeListing}
This section shows the code to trigger the applications of all transformations.
Each task is represented by a class containing related code.
Note that the classes contain some redundant code in order to allow the reader to concentrate on certain parts without hiding functionality by inheritance and call hierarchies.

\lstset{language=Java, 
        basicstyle=\small, % print whole listing small
        %keywordstyle=\color{black}\bfseries\underbar, % underlined bold black keywords
        identifierstyle=, % nothing happens
        commentstyle=\color{blue}, % blue comments
        stringstyle=\ttfamily, % typewriter type for strings
        showstringspaces=false, % no special string spaces
        numbers=left, numberstyle=\tiny, stepnumber=1, numbersep=5pt, % line numbers
        resetmargins=true, 
        breaklines=true,breakatwhitespace=true}

\begin{lstlisting}[caption={Abstract transformation base class},label={lst:atrafo}]
package de.unimarburg.swt.ttc2011;

import java.io.File;
import java.io.IOException;
import java.util.HashMap;

import org.eclipse.emf.common.util.URI;
import org.eclipse.emf.ecore.EObject;
import org.eclipse.emf.ecore.resource.Resource;
import org.eclipse.emf.ecore.resource.ResourceSet;
import org.eclipse.emf.ecore.resource.impl.ResourceSetImpl;
import org.eclipse.emf.ecore.xmi.XMIResource;
import org.eclipse.emf.ecore.xmi.impl.EcoreResourceFactoryImpl;
import org.eclipse.emf.ecore.xmi.impl.XMIResourceFactoryImpl;
import org.eclipse.emf.henshin.model.impl.HenshinPackageImpl;
import org.eclipse.emf.henshin.model.resource.HenshinResourceFactory;
import org.eclipse.emf.henshin.trace.impl.TracePackageImpl;

/**
 * This files contains some general methods and variables
 * used again and again. All other transformation related
 * classes inherit from that to use its given tool set as
 * loading and saving.
 * 
 * @author Stefan Jurack
 * 
 */
public abstract class ATrafo {
	
	public static final String MODELS_PATH = "HelloWorld/models/";
	public static final String HENSHIN_PATH = "HelloWorld/henshin/";
	public static final String OUTPUT_PATH = "HelloWorld/output/";
	
	/**
	 * The shared resource set for loading XMI files.
	 */
	protected final ResourceSet resourceSet = new ResourceSetImpl();
	
	/**
	 * To be overridden!
	 */
	public void start() {
		// Prerequisites
		initializeFactories();
		// ------
	}// start
	
	/**
	 * Loads the model and returns the root element.
	 * 
	 * @param modelFilePath
	 * @return
	 */
	protected EObject loadModel(String modelFilePath) {
		URI modelUri = URI.createFileURI(new File(modelFilePath)
				.getAbsolutePath());
		Resource modelResource = resourceSet.getResource(
				modelUri, true);
		if (modelResource.isLoaded()) {
			/*
			 * make sure that we work with a virgin instance
			 * of the model
			 */
			modelResource.unload();
			
			try {
				modelResource.load(null);
			} catch (IOException e) {
			}// try catch
		}// if
		return modelResource.getContents().get(0);
	}// loadModel
	
	/**
	 * Saves the content of the given <code>root</code>
	 * {@link EObject} to the specified file path.
	 * 
	 * @param filename
	 * @param root
	 */
	protected void saveModel(String filename, EObject root) {
		URI modelUri = URI.createFileURI(new File(OUTPUT_PATH
				+ filename).getAbsolutePath());
		Resource res = resourceSet.createResource(modelUri);
		HashMap<String, Object> opts = new HashMap<String, Object>();
		opts.put(XMIResource.OPTION_SCHEMA_LOCATION, true);
		try {
			res.getContents().add(root);
			res.save(opts);
		} catch (IOException e) {
			e.printStackTrace();
		}// try catch
	}// saveModel
	
	/**
	 * Initializes related factories and packages.
	 */
	protected void initializeFactories() {
		Resource.Factory.Registry.INSTANCE
				.getExtensionToFactoryMap().put("ecore",
						new EcoreResourceFactoryImpl());
		Resource.Factory.Registry.INSTANCE
				.getExtensionToFactoryMap().put("xmi",
						new XMIResourceFactoryImpl());
		Resource.Factory.Registry.INSTANCE
				.getExtensionToFactoryMap().put("henshin",
						new HenshinResourceFactory());
		
		HenshinPackageImpl.init();
		TracePackageImpl.init();
	}// initializeFactories
	
}// class
\end{lstlisting}
\begin{lstlisting}[caption={Starter for the transformations},label={lst:starter}]
package de.unimarburg.swt.ttc2011;

public class Starter {
	
	public static void main(String args[]) {
		ATrafo trafo;
		// -----
		
		trafo = new HelloWorldTrafo();
		trafo.start();
		// -----
		trafo = new CountMatchesTrafo();
		trafo.start();
		// -----
		trafo = new ReverseEdgesTrafo();
		trafo.start();
		// -----
		trafo = new SimpleMigrationsTrafo();
		trafo.start();
		// -----
		trafo = new DeleteNodeTrafo();
		trafo.start();
		// -----
		trafo = new InsertTransitiveEdgesTrafo();
		trafo.start();
		
	}// main
	
}// class
\end{lstlisting}
\begin{lstlisting}[caption={Hello World transformation class},label={lst:helloworld}]
package de.unimarburg.swt.ttc2011;

import org.eclipse.emf.ecore.EObject;
import org.eclipse.emf.henshin.common.util.EmfGraph;
import org.eclipse.emf.henshin.interpreter.EmfEngine;
import org.eclipse.emf.henshin.interpreter.UnitApplication;
import org.eclipse.emf.henshin.model.Rule;
import org.eclipse.emf.henshin.model.TransformationSystem;

/**
 * This class performs transformations related to the
 * HelloWorld case at the TTC 2011. <br>
 * Task 1
 * 
 * 
 * @author Stefan Jurack
 */
public final class HelloWorldTrafo extends ATrafo {
	
	public static final String HENSHIN_HELLOWORLD = HENSHIN_PATH
			+ "helloworld.henshin";
	public static final String HENSHIN_HELLOWORLD_EXT = HENSHIN_PATH
			+ "helloworldext.henshin";
	
	/**
     * 
     */
	@Override
	public void start() {
		super.start(); // prerequisites
		
		System.out.println("Task 1");
		
		performHelloWorld();
		performHelloWorldExt();
		
	}// start
	
	/**
	 * Creates an example "Hello World" instance.
	 */
	private void performHelloWorld() {
		
		TransformationSystem ts = (TransformationSystem) loadModel(HENSHIN_HELLOWORLD);
		EmfGraph emfGraph = new EmfGraph();
		EmfEngine engine = new EmfEngine(emfGraph);
		Rule rule = ts.findRuleByName("CreateSimple");
		UnitApplication unitApp = new UnitApplication(engine,
				rule);
		boolean success = unitApp.execute();
		if (success) {
			EObject result = (EObject) unitApp
					.getParameterValue("result");
			saveModel("HelloWorldSimple.xmi", result);
		} else {
			System.err
					.println("'Hello World' could not be applied!");
		}
	}// performHelloWorld
	
	/**
	 * Creates an example "Hello World" instance for the
	 * extended metamodel and subsequently performs a M2T
	 * transformation on that newly created model.
	 */
	private void performHelloWorldExt() {
		
		TransformationSystem ts = (TransformationSystem) loadModel(HENSHIN_HELLOWORLD_EXT);
		EmfGraph emfGraph = new EmfGraph();
		EmfEngine engine = new EmfEngine(emfGraph);
		Rule rule1 = ts.findRuleByName("CreateExtended");
		UnitApplication unitApp1 = new UnitApplication(engine,
				rule1);
		boolean success = unitApp1.execute();
		if (success) {
			EObject result = (EObject) unitApp1
					.getParameterValue("result");
			saveModel("HelloWorldExt.xmi", result);
		} else {
			System.err
					.println("'Hello World Ext' could not be applied!");
			return;
		}// if else
		
		/*
		 * We do now reuse the EmfGraph/Engine for the M2T
		 * task since then we already have an appropriate
		 * start graph at hand.
		 */

		Rule rule2 = ts.findRuleByName("M2T");
		UnitApplication unitApp2 = new UnitApplication(engine,
				rule2);
		success = unitApp2.execute();
		if (success) {
			EObject result = (EObject) unitApp2
					.getParameterValue("result");
			saveModel("HelloWorldM2T.xmi", result);
		} else {
			System.err
					.println("'Hello World M2T' could not be applied!");
		}// if else
		
	}// performHelloWorldExt
	
}// class
\end{lstlisting}
\begin{lstlisting}[caption={Count matches transformation class},label={lst:}]
package de.unimarburg.swt.ttc2011;

import org.eclipse.emf.ecore.EObject;
import org.eclipse.emf.henshin.common.util.EmfGraph;
import org.eclipse.emf.henshin.interpreter.EmfEngine;
import org.eclipse.emf.henshin.interpreter.UnitApplication;
import org.eclipse.emf.henshin.model.TransformationSystem;
import org.eclipse.emf.henshin.model.TransformationUnit;

/**
 * This class performs transformations related to the
 * HelloWorld task at the TTC 2011.<br>
 * Task 2
 * 
 * @author Stefan Jurack
 * 
 */
public class CountMatchesTrafo extends ATrafo {
	
	public static final String INSTANCE_GRAPH1 = MODELS_PATH
			+ "Graph1.xmi";
	
	public static final String HENSHIN_COUNT_MATCHES = HENSHIN_PATH
			+ "countMatches.henshin";
	
	/*
	 * (non-Javadoc)
	 * @see de.unimarburg.swt.ttc2011.ATrafo#start()
	 */
	@Override
	public void start() {
		super.start(); // prerequisites
		
		System.out.println("Task 2");
		
		performCountNumberOfNodes();
		performCountLoopingEdges();
		performCountIsolatedNodes();
		performCountCircles();
		performCountDanglingEdges();
		
	}// start
	
	private void performCountNumberOfNodes() {
		String unitName = "CountNodes";
		TransformationSystem ts = (TransformationSystem) loadModel(HENSHIN_COUNT_MATCHES);
		EObject rootObject = loadModel(INSTANCE_GRAPH1);
		EmfGraph emfGraph = new EmfGraph();
		emfGraph.addRoot(rootObject);
		EmfEngine engine = new EmfEngine(emfGraph);
		TransformationUnit unit = ts.findUnitByName(unitName);
		UnitApplication unitApp = new UnitApplication(engine,
				unit);
		boolean success = unitApp.execute();
		if (success) {
			EObject result = (EObject) unitApp
					.getParameterValue("counter");
			saveModel(unitName + "Result.xmi", result);
		} else {
			System.err.println("'" + unitName
					+ "' could not be applied!");
		}
	}// performCountNumberOfNodes
	
	private void performCountLoopingEdges() {
		String unitName = "CountLoopingEdges";
		TransformationSystem ts = (TransformationSystem) loadModel(HENSHIN_COUNT_MATCHES);
		EObject rootObject = loadModel(INSTANCE_GRAPH1);
		EmfGraph emfGraph = new EmfGraph();
		emfGraph.addRoot(rootObject);
		EmfEngine engine = new EmfEngine(emfGraph);
		TransformationUnit unit = ts.findUnitByName(unitName);
		UnitApplication unitApp = new UnitApplication(engine,
				unit);
		boolean success = unitApp.execute();
		if (success) {
			EObject result = (EObject) unitApp
					.getParameterValue("counter");
			saveModel(unitName + "Result.xmi", result);
		} else {
			System.err.println("'" + unitName
					+ "' could not be applied!");
		}
	}// performCountLoopingEdges
	
	private void performCountIsolatedNodes() {
		String unitName = "CountIsolatedNodes";
		TransformationSystem ts = (TransformationSystem) loadModel(HENSHIN_COUNT_MATCHES);
		EObject rootObject = loadModel(INSTANCE_GRAPH1);
		EmfGraph emfGraph = new EmfGraph();
		emfGraph.addRoot(rootObject);
		EmfEngine engine = new EmfEngine(emfGraph);
		TransformationUnit unit = ts.findUnitByName(unitName);
		UnitApplication unitApp = new UnitApplication(engine,
				unit);
		boolean success = unitApp.execute();
		if (success) {
			EObject result = (EObject) unitApp
					.getParameterValue("counter");
			saveModel(unitName + "Result.xmi", result);
		} else {
			System.err.println("'" + unitName
					+ "' could not be applied!");
		}
	}// performCountIsolatedNodes
	
	private void performCountCircles() {
		String unitName = "CountCircles";
		TransformationSystem ts = (TransformationSystem) loadModel(HENSHIN_COUNT_MATCHES);
		EObject rootObject = loadModel(INSTANCE_GRAPH1);
		EmfGraph emfGraph = new EmfGraph();
		emfGraph.addRoot(rootObject);
		EmfEngine engine = new EmfEngine(emfGraph);
		TransformationUnit unit = ts.findUnitByName(unitName);
		UnitApplication unitApp = new UnitApplication(engine,
				unit);
		boolean success = unitApp.execute();
		if (success) {
			EObject result = (EObject) unitApp
					.getParameterValue("counter");
			saveModel(unitName + "Result.xmi", result);
		} else {
			System.err.println("'" + unitName
					+ "' could not be applied!");
		}
	}// performCountCircles
	
	private void performCountDanglingEdges() {
		String unitName = "CountDanglingEdges";
		TransformationSystem ts = (TransformationSystem) loadModel(HENSHIN_COUNT_MATCHES);
		EObject rootObject = loadModel(INSTANCE_GRAPH1);
		EmfGraph emfGraph = new EmfGraph();
		emfGraph.addRoot(rootObject);
		EmfEngine engine = new EmfEngine(emfGraph);
		TransformationUnit unit = ts.findUnitByName(unitName);
		UnitApplication unitApp = new UnitApplication(engine,
				unit);
		boolean success = unitApp.execute();
		if (success) {
			EObject result = (EObject) unitApp
					.getParameterValue("counter");
			saveModel(unitName + "Result.xmi", result);
		} else {
			System.err.println("'" + unitName
					+ "' could not be applied!");
		}
	}// performCountDanglingEdges
	
}// class
\end{lstlisting}
\begin{lstlisting}[caption={Reverse edges transformation class},label={lst:reverseedges}]
package de.unimarburg.swt.ttc2011;

import org.eclipse.emf.ecore.EObject;
import org.eclipse.emf.henshin.common.util.EmfGraph;
import org.eclipse.emf.henshin.interpreter.EmfEngine;
import org.eclipse.emf.henshin.interpreter.UnitApplication;
import org.eclipse.emf.henshin.model.TransformationSystem;
import org.eclipse.emf.henshin.model.TransformationUnit;

/**
 * This class performs transformations related to the
 * HelloWorld task at the TTC 2011.<br>
 * Task 3
 * 
 * @author Stefan Jurack
 * 
 */
public class ReverseEdgesTrafo extends ATrafo {
	
	public static final String INSTANCE_GRAPH1 = MODELS_PATH
			+ "Graph1.xmi";
	
	public static final String HENSHIN_REVERSE_EDGES = HENSHIN_PATH
			+ "reverseEdges.henshin";
	
	@Override
	public void start() {
		super.start(); // prerequisites
		
		System.out.println("Task 3");
		
		performReverseEdges();
	}// start
	
	private void performReverseEdges() {
		String unitName = "ReverseEdges";
		TransformationSystem ts = (TransformationSystem) loadModel(HENSHIN_REVERSE_EDGES);
		EObject rootObject = loadModel(INSTANCE_GRAPH1);
		EmfGraph emfGraph = new EmfGraph();
		emfGraph.addRoot(rootObject);
		EmfEngine engine = new EmfEngine(emfGraph);
		TransformationUnit unit = ts.findUnitByName(unitName);
		UnitApplication unitApp = new UnitApplication(engine,
				unit);
		boolean success = unitApp.execute();
		if (success) {
			saveModel(unitName + "Result.xmi", rootObject);
		} else {
			System.err.println("'" + unitName
					+ "' could not be applied!");
		}// if else
	}// performReverseEdges
	
}// class
\end{lstlisting}
\begin{lstlisting}[caption={Simple migrations transformation class},label={lst:simplemigrations}]
package de.unimarburg.swt.ttc2011;

import org.eclipse.emf.ecore.EObject;
import org.eclipse.emf.henshin.common.util.EmfGraph;
import org.eclipse.emf.henshin.common.util.TransformationOptions;
import org.eclipse.emf.henshin.interpreter.EmfEngine;
import org.eclipse.emf.henshin.interpreter.UnitApplication;
import org.eclipse.emf.henshin.model.TransformationSystem;
import org.eclipse.emf.henshin.model.TransformationUnit;

/**
 * This class performs transformations related to the
 * HelloWorld task at the TTC 2011.<br>
 * Task 4
 * 
 * @author Stefan Jurack
 * 
 */
public class SimpleMigrationsTrafo extends ATrafo {
	
	public static final String INSTANCE_GRAPH1 = MODELS_PATH
			+ "Graph1.xmi";
	
	public static final String HENSHIN_SIMPLE_MIGRATION = HENSHIN_PATH
			+ "simpleMigration.henshin";
	public static final String HENSHIN_SIMPLE_MIGRATION_EXT = HENSHIN_PATH
			+ "simpleMigrationExt.henshin";
	
	@Override
	public void start() {
		super.start(); // prerequisites
		
		System.out.println("Task 4");
		
		performSimpleMigration();
		performSimpleMigrationExt();
	}// start
	
	private void performSimpleMigration() {
		String unitName = "SimpleMigration";
		TransformationSystem ts = (TransformationSystem) loadModel(HENSHIN_SIMPLE_MIGRATION);
		EObject rootObject = loadModel(INSTANCE_GRAPH1);
		EmfGraph emfGraph = new EmfGraph();
		emfGraph.addRoot(rootObject);
		EmfEngine engine = new EmfEngine(emfGraph);
		
		/*
		 * Note, that we set matching to be "non-injective"
		 * here. Thus, we can migrate Edges having equal
		 * source and target nodes. Alternatively, we just
		 * could have added one additional rule dealing with
		 * this special case.
		 */
		TransformationOptions options = new TransformationOptions();
		options.setInjective(false);
		engine.setOptions(options);
		
		TransformationUnit unit = ts.findUnitByName(unitName);
		UnitApplication unitApp = new UnitApplication(engine,
				unit);
		boolean success = unitApp.execute();
		if (success) {
			EObject result = (EObject) unitApp
					.getParameterValue("newGraph");
			saveModel(unitName + "Result.xmi", result);
		} else {
			System.err.println("'" + unitName
					+ "' could not be applied!");
		}
	}// performSimpleMigration
	
	private void performSimpleMigrationExt() {
		String unitName = "SimpleMigrationExt";
		TransformationSystem ts = (TransformationSystem) loadModel(HENSHIN_SIMPLE_MIGRATION_EXT);
		EObject rootObject = loadModel(INSTANCE_GRAPH1);
		EmfGraph emfGraph = new EmfGraph();
		emfGraph.addRoot(rootObject);
		EmfEngine engine = new EmfEngine(emfGraph);
		
		/*
		 * Note, that we set matching to be "non-injective"
		 * here. Thus, we can migrate Edges having equal
		 * source and target nodes. Alternatively, we just
		 * could have added one additional rule dealing with
		 * this special case.
		 */
		TransformationOptions options = new TransformationOptions();
		options.setInjective(false);
		engine.setOptions(options);
		
		TransformationUnit unit = ts.findUnitByName(unitName);
		UnitApplication unitApp = new UnitApplication(engine,
				unit);
		boolean success = unitApp.execute();
		if (success) {
			EObject result = (EObject) unitApp
					.getParameterValue("newGraph");
			saveModel(unitName + "Result.xmi", result);
		} else {
			System.err.println("'" + unitName
					+ "' could not be applied!");
		}
	}// performSimpleMigrationExt
	
}// class
\end{lstlisting}
\begin{lstlisting}[caption={Delete nodes transformation class},label={lst:deletenode}]
package de.unimarburg.swt.ttc2011;

import org.eclipse.emf.ecore.EObject;
import org.eclipse.emf.henshin.common.util.EmfGraph;
import org.eclipse.emf.henshin.common.util.TransformationOptions;
import org.eclipse.emf.henshin.interpreter.EmfEngine;
import org.eclipse.emf.henshin.interpreter.UnitApplication;
import org.eclipse.emf.henshin.model.TransformationSystem;
import org.eclipse.emf.henshin.model.TransformationUnit;

/**
 * This class performs transformations related to the
 * HelloWorld task at the TTC 2011.<br>
 * Task 5
 * 
 * @author Stefan Jurack
 * 
 */
public class DeleteNodeTrafo extends ATrafo {
	
	public static final String INSTANCE_GRAPH1 = MODELS_PATH
			+ "Graph1.xmi";
	
	public static final String HENSHIN_DELETE_NODE = HENSHIN_PATH
			+ "deleteNode.henshin";
	
	@Override
	public void start() {
		super.start(); // prerequisites
		
		System.out.println("Task 5");
		
		performDeleteNode();
		performDeleteNodeWithAllEdges();
	}// start
	
	private void performDeleteNode() {
		String unitName = "DeleteNodeN1Simple";
		TransformationSystem ts = (TransformationSystem) loadModel(HENSHIN_DELETE_NODE);
		EObject rootObject = loadModel(INSTANCE_GRAPH1);
		EmfGraph emfGraph = new EmfGraph();
		emfGraph.addRoot(rootObject);
		EmfEngine engine = new EmfEngine(emfGraph);
		TransformationUnit unit = ts.findUnitByName(unitName);
		UnitApplication unitApp = new UnitApplication(engine,
				unit);
		boolean success = unitApp.execute();
		if (success) {
			saveModel(unitName + "Result.xmi", rootObject);
		} else {
			System.err.println("'" + unitName
					+ "' could not be applied!");
		}// if else
	}// performDeleteNode
	
	private void performDeleteNodeWithAllEdges() {
		String unitName = "DeleteNodeN1WithAllEdges";
		TransformationSystem ts = (TransformationSystem) loadModel(HENSHIN_DELETE_NODE);
		EObject rootObject = loadModel(INSTANCE_GRAPH1);
		EmfGraph emfGraph = new EmfGraph();
		emfGraph.addRoot(rootObject);
		EmfEngine engine = new EmfEngine(emfGraph);
		
		/*
		 * Note that we use non-injective matching here.
		 * This is not required for "n1" as there are no
		 * cyclic edges, but in general this should be
		 * enabled for such cases (or alternatively
		 * considered during rule creation).
		 */
		TransformationOptions options = new TransformationOptions();
		options.setInjective(false);
		engine.setOptions(options);
		
		TransformationUnit unit = ts.findUnitByName(unitName);
		UnitApplication unitApp = new UnitApplication(engine,
				unit);
		boolean success = unitApp.execute();
		if (success) {
			saveModel(unitName + "Result.xmi", rootObject);
		} else {
			System.err.println("'" + unitName
					+ "' could not be applied!");
		}// if else
	}// performDeleteNodeWithAllEdges
	
}// class
\end{lstlisting}
\begin{lstlisting}[caption={Insert transitive edges transformation class},label={lst:inserttransitiveedges}]
package de.unimarburg.swt.ttc2011;

import org.eclipse.emf.ecore.EObject;
import org.eclipse.emf.henshin.common.util.EmfGraph;
import org.eclipse.emf.henshin.interpreter.EmfEngine;
import org.eclipse.emf.henshin.interpreter.UnitApplication;
import org.eclipse.emf.henshin.model.TransformationSystem;
import org.eclipse.emf.henshin.model.TransformationUnit;

/**
 * This class performs transformations related to the
 * HelloWorld task at the TTC 2011.<br>
 * Task 6
 * 
 * @author Stefan Jurack
 * 
 */
public class InsertTransitiveEdgesTrafo extends ATrafo {
	
	public static final String INSTANCE_GRAPH1 = MODELS_PATH
			+ "Graph1.xmi";
	
	public static final String HENSHIN_INSERT_EDGES = HENSHIN_PATH
			+ "insertTransitiveEdges.henshin";
	
	@Override
	public void start() {
		super.start(); // prerequisites
		
		System.out.println("Task 6");
		
		performInsertTransitiveEdges();
	}// start
	
	private void performInsertTransitiveEdges() {
		String unitName = "InsertTransitiveEdges";
		TransformationSystem ts = (TransformationSystem) loadModel(HENSHIN_INSERT_EDGES);
		EObject rootObject = loadModel(INSTANCE_GRAPH1);
		EmfGraph emfGraph = new EmfGraph();
		emfGraph.addRoot(rootObject);
		EmfEngine engine = new EmfEngine(emfGraph);
		TransformationUnit unit = ts.findUnitByName(unitName);
		UnitApplication unitApp = new UnitApplication(engine,
				unit);
		boolean success = unitApp.execute();
		if (success) {
			saveModel(unitName + "Result.xmi", rootObject);
		} else {
			System.err.println("'" + unitName
					+ "' could not be applied!");
		}
	}// performInsertTransitiveEdges
	
}// class
\end{lstlisting}

\end{appendix}

\end{document}